\documentclass[useAMS,usenatbib]{mn2e}
\usepackage{epsfig}

\DeclareMathVersion{bold}

\def\lsim{\,\lower2truept\hbox{${<\atop\hbox{\raise4truept\hbox{$\sim$}}}$}\,}
\def\gsim{\,\lower2truept\hbox{${>
\atop\hbox{\raise4truept\hbox{$\sim$}}}$}\,}

\title[Blind and non-blind source detections]{Blind and non-blind source
detection in WMAP 5-year maps}

\author[Massardi et al.]{
\parbox[t]{\textwidth}
{M.~Massardi$^{1,2}$\thanks{E-mail: massardi@sissa.it},
M.~L\'opez-Caniego$^{3}$, J.~Gonz\'alez-Nuevo$^{1}$,
D.~Herranz$^{4}$, G.~De~Zotti$^{5,1}$, and J.~L.~Sanz$^{4, 6}$}
\vspace*{8pt} \\
  $^{1}$ SISSA-I.S.A.S, via Beirut 4, I-34014 Trieste, Italy \\
  $^{2}$ Australia Telescope National Facility, CSIRO, PO Box 76 Epping, NSW
1710, Australia \\
  $^{3}$ Astrophysics Group, Cavendish Laboratory,
  J.J. Thomson Avenue, CB3 0E1, Cambridge, United Kingdom\\
  $^{4}$ Instituto de F\'\i{sica} de Cantabria (CSIC-UC), Avda. los
  Castros s/n, 39005 Santander, Spain \\
  $^{5}$ INAF-Ossevatorio Astronomico di Padova, vicolo
  dell'Osservatorio 5, I-35122 Padova, Italy\\
  $^{6}$ CNR Istituto di Scienza e Tecnologie dell'Informazione, I-56124, Pisa,
Italy}

\begin{document}

\maketitle

\begin{abstract}
We have analyzed the efficiency in source detection and flux density estimation
of \emph{blind} and \emph{non-blind} detection techniques exploiting the MHW2
filter applied to the Wilkinson Microwave Anisotropy Probe (WMAP) 5-year maps.
A comparison with the AT20G Bright Source Sample (Massardi et al.\ 2008), with
a completeness limit of 0.5 Jy and accurate flux measurements at 20 GHz, close
to the lowest frequency of WMAP maps, has allowed us to assess the completeness
and the reliability of the samples detected with the two approaches, as well as
the accuracy of flux and error estimates, and their variations across the sky.
The uncertainties on flux estimates given by our procedure turned out to be
about a factor of 2 lower than the rms differences with AT20G measurements,
consistent with the smoothing of the fluctuation field yielded by map
filtering. Flux estimates were found to be essentially unbiased except that,
close to the detection limit, a substantial fraction of fluxes are found to be
inflated by the contribution of underlying positive fluctuations. This is
consistent with expectations for the Eddington bias associated to the true
errors on flux density estimates. The blind and non-blind approaches are found
to be complementary: each of them allows the detection of sources missed by the
other. Combining results of the two methods on the WMAP 5-year maps we have
expanded the non-blindly generated New Extragalactic WMAP Point Source (NEWPS)
catalogue (L\'opez-Caniego et al.\ 2007) that was based on WMAP 3-year maps.
After having removed the probably spurious objects not identified with known
radio sources, the new version of the NEWPS catalogue, NEWPS\_5yr comprises 484
sources detected with a signal-to-noise ratio $\hbox{SNR} \ge 5$.
\end{abstract}


\section{Introduction} \label{sec:intro}

The best spectral region to study the Cosmic Microwave Background
(CMB) is the millimetric wavelength band, where the overall
contamination from foregrounds is at a minimum and the CMB black
body spectrum is close to its peak. An important byproduct of CMB
experiments is information on point sources, whose millimeter-wave
properties are poorly known. On the other hand, a careful
extraction of point sources from CMB maps is crucial since they
are the main foreground contaminant on small angular scales (less
than $\sim 30'$; De Zotti et al. 1999; Toffolatti et al. 1999).

The WMAP mission has produced the first all-sky surveys of extragalactic
sources at 23, 33, 41, 61 and 94~GHz. The analysis of first year data yielded a
sample of 208 extragalactic sources detected above a flux limit of $\sim
0.8$-1~Jy (Bennett et al. 2003), with an estimated completeness limit of $\sim
1.2\,$Jy at 23 GHz (Arg\"ueso et al. 2003; De Zotti et al. 2005). The sample
size has been steadily increasing as the WMAP survey successfully progressed:
323 sources were found in the 3-yr maps (Hinshaw et al. 2007; we will refer to
this sample as WMAP\_3yr), and 390 in the 5-yr maps (Wright et al. 2008;
WMAP\_5yr sample). The approach used by the WMAP team for source extraction can
be summarized as follows. The temperature map pixels were first weighted by
$N_{\rm obs}^{1/2}$, $N_{\rm obs}$ being the number of independent observations
per pixel, and then filtered in harmonic space by the global matched filter
$b_l/(b_l^2C_l^{\mathrm{CMB}}+C_l^{\mathrm{noise}})$, where $b_l$ is the
transfer function of the WMAP beam response, $C_l^{\mathrm{CMB}}$ is the CMB
angular power spectrum, and $C_l^{\mathrm{noise}}$ is the noise power spectrum.
Peaks with signal-to-noise ratio ($\hbox{SNR}$) greater than $5$ (note that the
`noise' here is the global rms fluctuation in regions outside the processing
mask) in any band were interpreted as source detections. The peaks are fitted
in real space, i.e. in the unfiltered maps, to a Gaussian profile plus a planar
baseline to estimate the flux densities. The flux densities in the other
channels are given if their $\hbox{SNR} > 2$ and the source width falls within
a factor of two of the true beam width.

Several other attempts to improve the source detection have been
presented. Nie \& Zhang (2007) applied cross-correlation
techniques to clean the WMAP first-year residual maps and identify
foreground residuals which have been associated to radio sources:
they detected 101 sources of which 26 where not in the WMAP 1-year
catalogue (25 of them do not appear even in the WMAP\_3yr
catalogue, but 5 of them are in the LMC region). Chen \& Wright
(2008), combining the 61 and 94 GHz WMAP temperature maps to
cancel the `noise' due to the CMB anisotropy signal, found 31
sources in the first year maps and 64 in the 3-year co-added maps,
of which 21 are not in WMAP\_3yr. The same $V-W$ internal linear
combination technique was used by Wright et al. (2008) to find 99
sources in the region with $|b| > 10^\circ$, 64 of which are in
WMAP\_5yr, 17 can be identified with known sources, 17 are in
complex Galactic emission regions, and 1 is unidentified.

In previous works, we have used a non-blind approach (see
L\'opez-Caniego et al.\ 2007, Gonz\'alez-Nuevo et al.\ 2008,
hereafter LC07 and GN08, respectively) exploiting the MHW2 filter
(Gonz\'alez-Nuevo et al.\ 2006) to obtain estimates of (or upper
limits to) the flux densities at the WMAP frequencies for a complete
all-sky sample of 2491 sources with $|b|>5^\circ$, brighter than 500
mJy at 5 GHz in the PMN (Griffith et al.\ 1993, 1995; Wright et al.\
1994, 1996) or in the GB6 (Gregory et al.\ 1996) catalogs, or at 1.4
or 0.84 GHz in regions not covered by 5 GHz surveys but covered by
either the NVSS (Condon et al.\ 1998) or the SUMSS (Mauch et al.\
2003, 2007). This work yielded $5\sigma$ detections of 380
extragalactic sources in the WMAP 3-yr maps, including 98 sources
not present in the WMAP\_3yr catalog. The results were organized in
the NEWPS (New Extragalactic WMAP Point Source) catalog (since it is
based on 3-year maps we will hereafter add the suffix `3yr' to the
name).

In this paper we extend the analysis to the WMAP 5-yr data, carrying
out both a `blind' and a `non-blind' source search using the same
filter (MHW2), that was shown by L\'opez-Caniego et al.\ (2006) to
be essentially as efficient as the matched filter, and easier to
use. A particularly delicate issue that we will address is the
estimate of the `noise' to be used to derive the $\hbox{SNR}$ and
hence the nominal flux limit for source detection, in the presence
of a highly inhomogeneous  fluctuation field. This analysis is
important also in view of defining the optimal source extraction
strategy for the Planck mission.

The strong inhomogeneity and non-Gaussianity of the fluctuation field, which in
the best CMB experiments is dominated not by the instrumental noise but by
sources below the detection limit and by small-scale structure in the Galactic
emission, is a serious hindrance for source detection techniques. Since the
statistical properties of such fluctuation field are poorly known, the
reliability of source detections and the real uncertainties on flux estimates
are difficult to quantify, even in the case of relatively high $\hbox{SNR}$'s.
It is well known (Eddington 1913; Murdoch et al. 1973; Hogg \& Turner 1998)
that the skewness of the distribution of Poisson fluctuations due to unresolved
sources may strongly bias flux estimates with $\hbox{SNR} < 5$, and the effect
is larger for steeper source counts. Source clustering and small-scale
structure of the Galactic emission may substantially worsen the problem for low
resolution experiments, such as those aimed at mapping the CMB. Simulations of
Planck observations (Leach et al. 2008) show that both the fraction of spurious
detections and the incompleteness level may be of several percent, even at flux
limits corresponding to $\hbox{SNR} \ge 5$.

Fortunately the Bright Source Sample (BSS; Massardi et al. 2008, M08
hereafter), complete down to $S_{20 \rm GHz}=0.5\,$Jy for $\delta < -15^\circ$,
obtained from the Australia Telescope 20~GHz (AT20G) survey, offers the
opportunity of an empirical assessment of the completeness and of the
reliability of samples extracted from the WMAP 23 GHz map in the same area.
Follow-up observations at 20 GHz have yielded accurate flux measurements,
allowing us to determine the accuracy of flux and error estimates at the nearby
WMAP frequency of 23 GHz. Spectral information is also available, thanks to
measurements at 8.6, 4.85 GHz, and, for a sub-sample, at 95~GHz (Sadler et al.
2008). The lessons learned from the comparison of the results of the analysis
of 23 GHz maps with the AT20G data provided an useful guidance for the
investigation of WMAP all-sky data also at the other WMAP frequencies. We have
limited our study to the first 4 WMAP channels, leaving aside the 94 GHz
channel because of the normalization problems discussed by LC07 and GN08.

This paper is organized as follows. In \S\,\ref{sec:technique} we
describe our approach to source detection. The reliability and
completeness of the samples of detected sources, and the quality
of flux density estimations are discussed, using the BSS data as a
benchmark, in \S\,\ref{sec:AT20G}. In \S\,\ref{sec:newps5yr} we
extend the analysis to the whole $|b|>5^\circ$ WMAP maps and
describe the properties of the NEWPS\_5yr catalogue. Finally, in
\S\,\ref{sec:Conclusions}, we summarize and briefly discuss our
main conclusions.

\begin{table*}
  \centering
\begin{tabular}{lccc}
\hline
Sample ID                     & NEWPS\_3yr                   & WMAP 5-yr NB
\\
\hline
Method                        & non-blind                    & non-blind
\\
Input positions (for NB)      &5 GHz catalogues              &
NEPWS\_3yr\_3$\sigma$                   \\
Sky coverage                  & All sky with $|b|>5^\circ$   &  All sky with
$|b|>5^\circ$              \\
$\hbox{SNR}>3$ detections          & $[759,\ 564,\ 535,\ 365]$    & $[712,\
585,\ 537,\ 312]$             \\
$\hbox{SNR}>5$ detections          & $[349,\ 223,\ 217,\ 135]$    & $[366,\
262,\ 246,\ 122]$              \\
median flux density error (mJy)       & $[182,\ 219,\ 214,\ 251]$
& $[168,\
207,\ 196,\ 249]$             \\
min flux density at $\hbox{SNR}>5$ (mJy)  & $[712,\ 995,\ 861,\ 995]$    &
$[754,\ 888,\ 861,\ 950]$        \\
\hline Sample ID                    & WMAP 5-yr SB        & WMAP
5-yr IB
\\
\hline Method                       & simple blind &  iterative
blind
\\
Sky coverage                 & All sky with $|b|>5^\circ$   &  All sky with
$|b|>5^\circ$    \\
$\hbox{SNR}>3$ detections         & $[1826,\ 2279,\ 3001,\ 3441]$& $[1302,\
1345,\ 1575,\ 1308]$  \\
$\hbox{SNR}>5$ detections         & $[454,\ 304,\ 285,\ 155]$    & $[399,\
279,\ 265,\ 143]$      \\
median flux density error (mJy)      & $[167,\ 206,\ 196,\ 249]$    & $[168,\
206,\ 194,\ 247]$      \\
min flux density at $\hbox{SNR}>5$ (mJy) & $[695,\ 874,\ 831,\ 1082]$   &
$[744,\ 876,\ 854,\ 963]$  \\
 \hline
\end{tabular}
\caption{Summary of the main properties of the blind and non-blind
(NB) samples discussed in this paper. Values in the square
brackets refer respectively to $[23,\ 33,\ 41,\ 61]$~GHz. Note
that we have investigated 2 different blind approaches: a `simple'
blind (SB) and a `iterative' blind (IB); details are in the text.}
\label{tab:survey_summary}
\end{table*}

\section{Detection techniques} \label{sec:technique}

\begin{figure*}
\includegraphics[width=16cm]{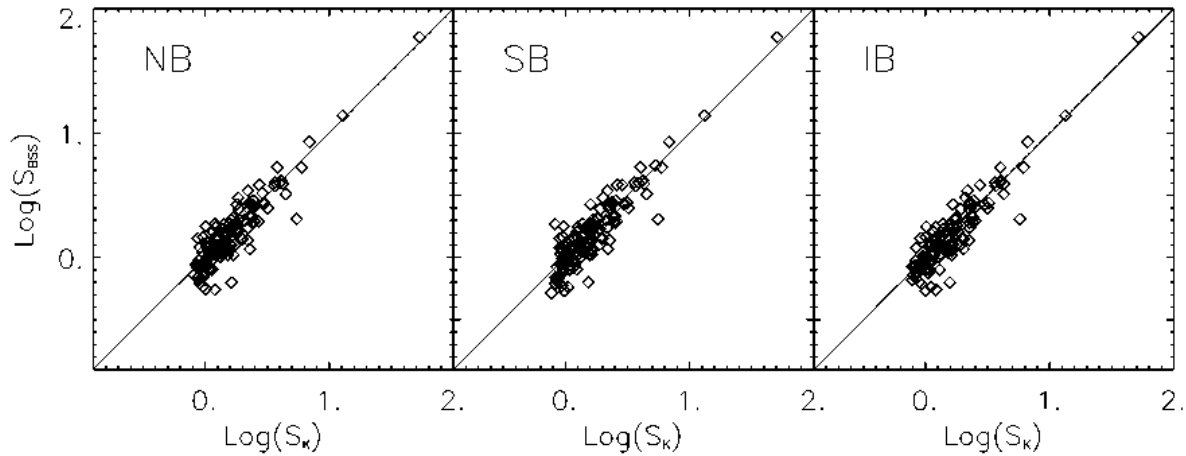}\\
\vspace{0.5cm} \caption{Comparison of flux densities estimated from WMAP maps
using each of our 3 methods with the AT20G ones. The agreement is good except
for the systematic offset at faint flux densities (see text for further
discussion).} \label{fig:atvsall}
\end{figure*}

The MHW2, that we use in the present study, is the second member
of the Mexican Hat Wavelet filter family (Gonz\'alez-Nuevo et al.
2006). It is obtained by analytically applying the Laplacian
operator twice on the 2D Gaussian function. It operates locally
removing simultaneously the large scale variations originated in
the diffuse Galactic foregrounds as well as the small scale noise.
The scale at which MHW2 operates can be easily optimized so that
the signal-to-noise ratio (SNR) of the sources is maximized. This
scale is obtained numerically in an easy way for any given sky
patch. The optimal scale is approximately equal to the
beamwidth and generally varies by no more than 10-15 percent
except in highly contaminated regions where the variation may be
of up to a factor of 2. A comparison of the shapes in the Fourier domain of some matched filters with the corresponding MHW2 at the optimal scale can be found in Lopez-Caniego et al. (2006).

After filtering, the flux is estimated at the position of the
maxima. The wavelet can be normalized in such a way that the
intensity value at the maxima is equal to the flux of the source.
This estimation of the flux is, on average, unbiased. The
normalisation of the wavelet is very sensitive to the assumed
profile of the signal. In LC07 it was shown how it is possible to
go beyond the Gaussian approximation, using the symmetrized radial
beam profiles provided by WMAP. In this work we have used the
updated 5 year beam profiles throughout the analysis.

In the {\it simple blind} (SB) approach we look for objects above
a given $\hbox{SNR}$ anywhere in the patch. In the {\it non-blind}
(NB) approach, whereby we are looking for WMAP sources at the
positions of previously known sources, the patch is chosen so that
the source position is right at the center of the patch, and we
measure the $\hbox{SNR}$ there.  Finally, the {\it iterative
blind} (IB) approach consists in producing, for each source
detected with the blind approach, new patches centered at the
source positions, and in re-estimating the $\hbox{SNR}$s.

For this work we have used an end-to-end code that reads in an
input parameter file containing the specific characteristics of
the maps to be studied, reads in the input map in FITS format,
extracts the patches to be analyzed using the tangential plane
approximation, finds for each patch the optimal scale of the
wavelet, filters each of them with the MHW2 code, produces a
list of detections above a given $\hbox{SNR}$, converts the
positions of the detected objects from the tangent plane to the
sphere and, finally, combines the detections into a single output
file.

In the input parameter file we specify how to obtain the patches
needed for the analysis. In the general case, the code divides and
projects the sky into a sufficient number of square patches such
that the whole sky is not only fully covered, but also there is a
sufficient amount of overlap among the patches to allow cuts of
the borders of the image, if needed. The size of the patches in
the sky, the pixel size and the amount of overlap among patches
are specified in the parameter file. We have used flat projected
patches of $14.6^\circ \times 14.6^\circ$, each containing $128
\times 128$ pixels. The  pixel area is $6.87' \times 6.87'$,
corresponding to the HEALPix resolution parameter $\hbox{N}_{\rm
side}=512$. The patch making routine is part of the CPACK
library\footnote{http://astro.ic.ac.uk/$\sim$mortlock/cpack/}.
There is also the option of inputting the list of coordinates of
the centers of the patches, corresponding to the known positions
of the sources in the cases of the non-blind and of the iterative
blind approaches. In the following subsections we will describe in
detail how the algorithms work for each approach.

The comparison of fluxes determined from the WMAP 5-year maps with
those listed in the NEWPS\_3yr catalogue shows good agreement if
the same calibration, described in GN08, is applied. In fact, the
correction factors to the effective beam areas calculated using
the symmetrized beam profiles given by the WMAP team have not
changed significantly. Such correction factors are $[1.05, 1.086,
1.136, 1.15]$  at $[23, 33, 41, 61]$~GHz, respectively.

\subsection{Simple blind approach} \label{sec:blind_meth}

The program reads in the input parameter file and the map in FITS
format and calculates the number of flat patches to be extracted
and the coordinates of their centers. For our choice of the input
parameters ($14.6^\circ \times 14.6^\circ$ patches with $3^\circ$
overlap) the program extracts 371 flat patches. Next, the code
loops over each of them, finding the optimal scale, filtering the
maps with the MHW2 at such optimal scale and detecting objects
with $\hbox{SNR} \ge 3$. For each patch a temporary catalogue is
obtained, and for each object, the flux at the position of the
corresponding peak is estimated. Finally, the temporary catalogues
are combined into a final one, removing duplications (in the case
of multiple detections of the same source we select the one with
the brightest flux, that normally corresponds to the most accurate
position).

The rms of the map is obtained via a three step process. First, in
order to avoid border effects after filtering, a 15 pixel border
around the maps is flagged. Second, all the maxima in the image
are identified and a histogram of their values is obtained. Then,
the 5 per cent of the brightest maxima are masked, flagging the
pixels within a 2 FWHM radius from the position of the maxima.
Finally, the rms of the map is calculated excluding the flagged
pixels.

\subsection{Non-blind approach}\label{sec:nonblind_meth}

In the non-blind approach the patches to be analyzed are centered
at the positions of the objects we want to investigate. Since the
position of the source is already known, the goal is to get a good
characterization of the noise rms level in its vicinity. The
algorithm goes as for the blind approach, with the following
differences: i) we have an additional input file, containing the
coordinates of the objects; ii) we look for maxima within a circle
around the patch center, with 1 FWHM radius; iii) the rms
fluctuation level is estimated taking into account only a corona
around the patch center, with inner radius of 1 FWHM and an outer
radius of 3 FWHM.

In practice, the amplitude of the central maximum (if any) gives
an estimate of the source flux, and to compute the rms noise we
apply the flagging of pixels at the border, the search of maxima,
and the flagging of the 5 per cent brightest, only to the corona.
In this way, we try to get a more accurate estimate of the noise
in the vicinity of the object of interest, avoiding the
contamination by other bright nearby objects.

The application of this approach builds on the work by LC07 who
have looked for signals in WMAP 3-year maps at the positions of
2491 sources forming a complete sample mostly selected at 5 GHz.
They detected 369 of these sources with $\hbox{SNR}\ge 5$ in at
least one WMAP channel. The detection efficiency is therefore of
only 14.8\%. The lower noise level in WMAP 5-year data can allow
the $\hbox{SNR}\ge 5$ detection of some, but not many, fainter
sources. Thus, we limited our non-blind search to the 933 sources
in LC07 associated to $\hbox{SNR}\ge 3$ peaks in the 3-yr maps.

As discussed below, the AT20G Bright Source Sample (BSS), complete
to $S_{20\rm GHz}=0.5\,$Jy and covering about $1.5\times
10^4\,\hbox{deg}^2$, is particularly useful to test the
performances of our detection algorithms. Our non-blind approach
was first applied to this sample.

\subsection{Iterative blind approach}\label{sec:combimet}

The iterative blind method proceeds in two steps. The first step
follows the procedure described in \S\,\ref{sec:blind_meth} and
produces as output a list of coordinates that are fed to the
non-blind scheme of \S\,\ref{sec:nonblind_meth}. In this way we
hoped to combine the potential of the blind detection with the
advantages of the local noise estimation of the non-blind method.

\subsection{Sampling of the fluctuation field}\label{sec:sampling}
The choice of the region to be used for estimating the noise level is the best compromise between the conflicting requirements of sufficient statistics and not being affected by variations of the background far from the source position. In fact, we are interested in getting the best estimate of fluctuations at the source position.

This issue may be critical in the cases of the non-blind or iterative blind approaches, since the region where the fluctuation field is sampled is rather limited. However, we have typically 30 independent areas within the sampling region. Using the formalism given by Danese et al. (1980) we find that, for a Gaussian fluctuation field, the uncertainty on the rms fluctuation is  $[+ 16\%, -12\%]$. In the case of the simple blind approach we have about 180 independent areas in each region and the uncertainty drops to $\simeq \pm 5\%$.

\begin{figure}
\includegraphics[width=7cm,angle=90]{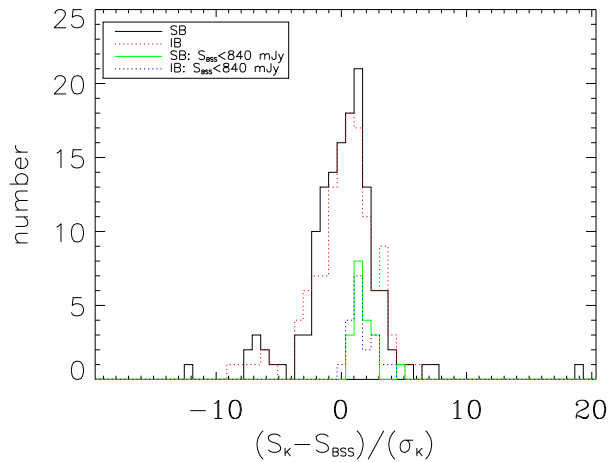}\\
\caption{Distribution of the ratio of `true' to estimated errors.
$S_{\rm BSS}$ is the ATCA flux density, measured with high SNR,
that we assume to be the `true' value. $S_{\rm K}$ and
$\sigma_{\rm K}$ are our flux and error estimates from the WMAP
K-band map with the SB and IB methods (see inset). The histograms
labeled $S_{\rm BSS}< 840\,$mJy in the inset include only the
faint sources whose $S_{\rm K}$ is systematically higher than
$S_{\rm BSS}$ (see Fig.~\protect\ref{fig:atvsall}).}
\label{fig:ratio}
\end{figure}

\section{Blind vs non-blind detection: comparison with AT20G
data}\label{sec:AT20G}

The prior knowledge of source coordinates has the obvious
advantage that source detection algorithms need to determine only
one parameter, i.e. the source flux, while blind detection must
deal also with the 2 additional parameters defining the source
position, and are exposed to be misled by source blending or
small-scale structure in the Galactic emission.

On the other hand, catalogues obtained from a non-blind approach
are liable to various possible sources of incompleteness ensuing
from the selection that has produced the input catalogue. The
latter may have been generated by a survey at a different
frequency (generally lower than WMAP's), with different angular
resolution (generally much higher than WMAP's), carried out at a
different epoch. A lower frequency survey may easily miss sources
with strongly inverted (i.e. increasing with increasing frequency)
spectra. High angular resolution observations (especially the
interferometric ones) are insensitive to extended sources, or may
pick up only their compact spots, while the sources may be much
brighter at the WMAP resolution. Observations at different epochs
may catch variable sources in different stages, so that a source
that is too faint to be included in the input catalogue may be
detected by WMAP, and vice versa.

The AT20G Bright Source Sample (BSS; M08) minimizes the problems
mentioned above: i) it has been selected at 20 GHz, i.e. at a
frequency close to that of the WMAP K-band channel; ii) the survey
has been carried out from 2004 to 2007, i.e. in a period
overlapping that of WMAP 5-year maps (obtained averaging over the
data collected in 2001-2006). As pointed out by Sadler et al.
(2006), on a 1–-2 yr time-scale, the general level of variability
at 20 GHz has a median value of 6.9 percent, which is low compared to our uncertainties on flux density estimates. The
only completeness problem of the BSS for our analysis is related
to the size ($2.4'$) of the 20 GHz ATCA primary beam. The ensuing
incomplete sampling of extended sources will be discussed in the
following. Because of its properties, the AT20G BSS constitutes an
excellent benchmark against which we may test the performances of
blind and non-blind detection techniques applied to the WMAP 23
GHz maps.

Thus, first of all we have performed the Simple Blind (SB) and
Iterative Blind (IB) searches on the WMAP 5-yr 23~GHz maps and
analyzed the results over the area of the AT20G BSS ($\delta
<-15^\circ$), cutting out the Galactic plane region
($|b|<5^\circ$). This cut removes 26 of the 320 BSS sources. Of
the remaining 294 sources, 124 have $S_{\rm BSS}>1\,$Jy. Next we
repeated the search non-blindly, on patches centered at the BSS
source positions on the WMAP K-band map. The non-blind technique
detected 125 sources (96 with $S_{20\rm GHz}>1\,$Jy) with
$\hbox{SNR}>5$; the mean flux density error is of 212 mJy and the
minimum detected flux density is of 767 mJy.

The association of peaks in WMAP maps with BSS sources was made adopting a
search radius of $21.35'$, i.e. equal to $\sigma=\hbox{FWHM}/2\sqrt{2\ln 2}$
for $\hbox{FWHM}=50.277'$. The position of a detection is given by the
coordinates of the pixel where a local maximum is found. The median of the
distances of the SB detections from the real positions of the sources (given by
the AT20G BSS) is $3.7'$ (for the iterative blind it is $3.3'$), so the
positions are typically correct within a pixel for most ($\sim 83\%$) of the
objects (the pixel size for the WMAP maps is $6.87'$).

The simple blind search recovers 140  BSS sources: 114 are in common
with the non-blindly detected sample, but 26 BSS sources have been detected
only blindly; 14 BSS sources with $S_{\rm BSS}>1\,$Jy remain undetected at
$\hbox{SNR}>5$ level. The iterative blind search, on the other hand, recovers
128 BSS sources: 115 are in common with the non-blindly detected sample, and 13
BSS sources have been detected only blindly; 27 sources remain undetected. Flux
estimations are consistent with the BSS measurements, at least for the
brightest objects. All the sources with $S_{\rm BSS}>1\,$Jy undetected by
either method show up as local maxima with $\hbox{SNR}<5$; most of their fluxes
are underestimated, probably because they lie over a negative fluctuation peak.
Furthermore the SB and IB searches yielded, respectively, 41 and 30 detections
of objects that are not in the BSS.

\begin{figure*}
\includegraphics[width=18cm]{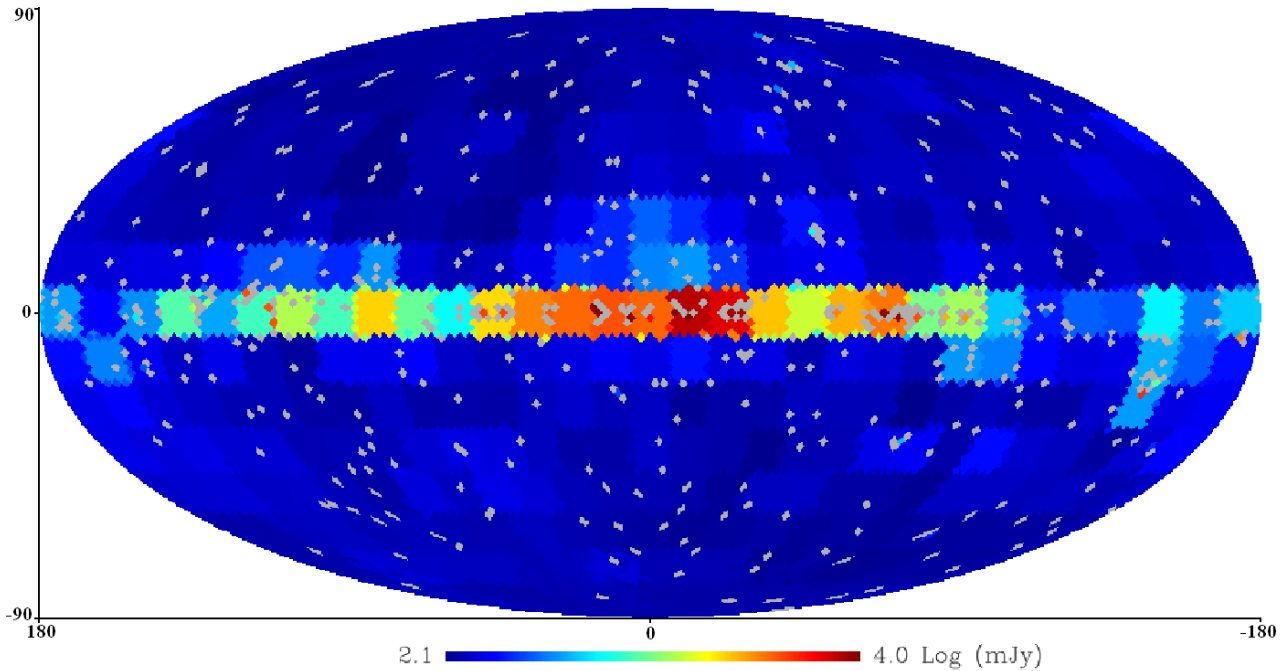}\\
\caption{Map (Mollweide projection in Galactic coordinates) of $\sigma_{\rm
pixel}$ for the SB approach at 23 GHz. The pixel area is of $\simeq
3.36\,\hbox{deg}^2$ (HEALpix pixelization with $\hbox{N}_{\rm side}=32$). The
patches and the $3^\circ$ overlaps among detection patches are discernible.
Red regions correspond to higher values of noise, darker blue region
correspond to lower values (see the colour scale at the bottom of the figure).
Grey dots are pixels where no sources are found. A color version of the figure
is available on-line.} \label{fig:noisemap_23}
\end{figure*}

\subsection{Accuracy of flux density and error estimates}\label{sec:acc}

A comparison of the flux densities derived from the WMAP maps with
those measured by the AT20G survey is shown in
Fig.~\ref{fig:atvsall}. The agreement is generally good, except
for the faintest levels ($S_{\rm BSS}< 840\,$mJy) where the WMAP
fluxes are systematically higher. Since there are no indications
that the faint sources are extended and may therefore be resolved
by the AT20G, the discrepancy is likely due to a swelling of the
peak at WMAP resolution when the sources happen to be on top of
large positive fluctuations due to other components (noise,
Galactic emission, CMB). There is no obvious way to identify
sources affected by this problem using WMAP data only, and this is
another instance of the importance of complementary, higher
resolution data. Such sources are not found in particularly
contaminated regions; on the contrary, the associated noise values
are generally rather low.


The comparison of AT20G flux densities, measured with high SNRs,
that we assume to be the `true' values, with the flux and error
estimates from WMAP maps allow us to assess also the reliability
of error estimates. The results are illustrated by
Fig.~\ref{fig:ratio}. After having removed the sources with
$S_{\rm BSS}< 840\,$mJy, whose fluxes are systematically
overestimated, the median $(S_{\rm K}-S_{\rm BSS})$ is 124 mJy for
the simple blind and 41 mJy for the iterative blind respectively.
Since the distribution is not too different from a
Gaussian, the error on the median can be estimated as (Arkin \&
Colton 1970) $\sigma_{\rm med}=1.2533/[\sum_i (1/\sigma_{\rm
K,i})^2]^{1/2}=18\,$mJy. An average $\langle(S_{\rm K}-S_{\rm
BSS})\rangle \simeq 40\,$mJy is expected for an average spectral
index $\alpha=0.3$ ($S_\nu \propto \nu^{-\alpha}$). The
simple blind method thus slightly overestimates (by $\lsim 10\%$)
the K-band fluxes, while in the case of the iterative blind method
the mean difference with BSS fluxes can be entirely attributed to
the small difference in frequency.

After correcting for the small offset, the standard
deviation of $(S_{\rm K}-S_{\rm BSS})/\sigma_{\rm K}$ is 2.4 for
the SB and 2.0 for the IB sample; more than 93\% of the sources
lie within 3 standard deviations from the mean (both including and
removing the sources with $S_{\rm BSS}< 840\,$mJy). The fact that
the rms differences between the BSS fluxes, $S_{\rm BSS}$, and our
estimates from WMAP K-band maps, $S_{\rm K}$, are about twice the
average $\sigma_{\rm K}$ is not surprising. As pointed out by
LC07, by applying the optimum filters to WMAP temperature maps we
get an average amplification of the SNR, or equivalently, a
damping of the fluctuation level by a substantial factor. In other
words, the fluctuation level in the original map
due to a combination of anisotropies in the CMB and
foreground emissions plus noise, that determines the true
uncertainty on the flux estimate, is substantially higher than
that in the filtered map, used to estimate $\sigma_{\rm K}$. The
ratio of the two noise levels is a measure of the detection
efficiency of the adopted algorithm.

\subsection{Reliability of detections}

As mentioned above, the SB and the IB approaches yielded a number of
detections without a BSS counterpart. To check whether these sources are real
we have looked for counterparts in the NASA Extragalactic Database
(NED\footnote{http://nedwww.ipac.caltech.edu/}). Of the 41 non-BSS
detections of the SB search, 15 were found to be extragalactic and 9  Galactic
sources, while of the 30 IB non-BSS sources 12 were found to be extragalactic,
and 4 Galactic. The BSS is biased against Galactic sources, mostly because
they are generally extended; in fact no known Galactic source is included in
it. It has also completely missed the very extended extragalactic source Fornax
A (detected in WMAP maps). All the other extragalactic sources that were
detected by our blind techniques and have counterparts in the NED were also
detected by the AT20G survey, but below the BSS flux density threshold. Since
accurate estimates of the total flux density of the known extended sources in
the area (except for Fornax A) have been obtained with ATCA observations in the
mosaic mode, the discrepancy cannot be attributed to resolution effects, and in
fact there is no indication that the sources in question are extended. We
therefore conclude that the K-band fluxes are overestimated, probably because
these sources happen to be on top of positive fluctuations of noise and/or
Galactic and/or CMB signals within the WMAP beam.

\begin{figure}
\includegraphics[width=8cm]{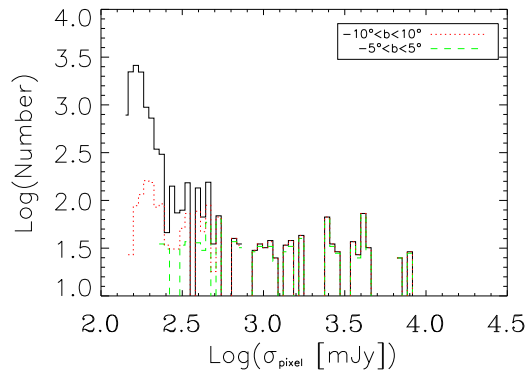}\\
\caption{Distribution of the values of $\sigma_{\rm pixel}$
(pixels size of $\simeq 3.36\,\hbox{deg}^2$) over the whole sky
(solid line), the region within $|b|<10^\circ$ (dotted line), and
the region within $|b|<5^\circ$ (dashed
line).}\label{fig:distr_noisemap_23}
\end{figure}

The 17 SB objects (or the 14 IB ones) that do not have a
consistent counterpart in the NED may be knots in the Galactic
emission, as suggested by the fact that 14 (or 9 for the
IB) of these sources are at $|b|<20^\circ$. To better
characterize the sky regions more liable to the occurrence of
spurious detections we have produced a 23 GHz noise map
(Fig.~\ref{fig:noisemap_23}) with pixels size of $\simeq
3.36\,\hbox{deg}^2$, corresponding to the HEALPix $\hbox{N}_{\rm
side}=32$ (the size of patches discussed above corresponds to
$\hbox{N}_{\rm side}=4$). Figure~\ref{fig:distr_noisemap_23} shows
that a $\pm10^\circ$ Galactic cut removes almost all the most
contaminated pixels, but also some clean regions. The $\pm5^\circ$
Galactic cut that we have used so far seems to be a better
compromise between removing very contaminated regions and saving
clean ones. However, the selection could be improved by selecting
a mask for contaminated regions exploiting the information given
by the noise map itself.

\begin{figure}
\includegraphics[width=6cm,height=9cm, angle=90]{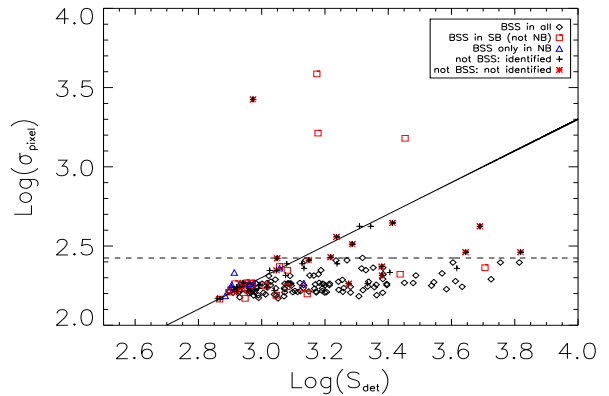}\\
\caption{Noise at the source position versus flux density at 23
GHz estimated with the SB approach. The dashed line corresponds to
1.5 times the median noise for the pixels at $|b|>5^\circ$. The
solid line corresponds to $S_{\rm K} = 5\sigma_{\rm pixel}$. 11
sources have $S_{\rm K} < 5\sigma_{\rm pixel}$ and correspond to
$\hbox{SNR}>5$ detections in highly contaminated pixels
($\sigma_{\rm pixel}\gg \sigma_{\rm K}$; remember that
$\sigma_{\rm K}$ is computed over a much larger area than
$\sigma_{\rm pixel}$). }\label{fig:flux_noise}
\end{figure}

As illustrated by Fig.~\ref{fig:flux_noise}, most (but not all) of the objects
that do not have a consistent counterpart in the NED lie in regions where the
noise level is relatively high. Dropping areas with $\sigma_{\rm pixel}\ge
1.5\sigma_{\rm median} = 253\,$mJy at 23 GHz, where $\sigma_{\rm
median}=169\,$mJy is the median noise level for all pixels at $|b|>5^\circ$,
removes 17 SB sources and 11 IB ones; 9 of the SB and 5 of the
IB sources removed are doubtful. We thus approximately halve the
number of possibly spurious SB sources at the cost of losing $\simeq 7\%$ of
the sky region with $|b|>5^\circ$ covered by the AT20G BSS (the remaining area
amounts to 3.77 sr). This criterion is a good trade-off between completeness
and reliability of the sample, and we will adopt it for the all-sky analysis.
As for the reliability, only 8 SB sources and 9 IB sources detected in
regions with $\sigma_{\rm pixel}< 1.5\sigma_{\rm median}$ do not have a
consistent low-frequency counterpart. If they are all spurious, the sample
reliability is 95.5\% for the SB and 94.3\% for the IB.

\subsection{Completeness}\label{completeness}

The inhomogeneity of the fluctuation field translates into a spatially varying
effective depth of the survey. Correspondingly, the effective area to be used
to derive the source counts decreases with decreasing flux limit. This is
illustrated by Fig.~\ref{fig:noisemap_23} showing the map of the noise within
the $\simeq 3.36\,\hbox{deg}^2$ pixels ($\sigma_{\rm pixel}$). Note that
$\sigma_{\rm pixel}$ is approximately the same in all the pixels within a
detection patch and vary on the edge of it because of the overlap among
patches. The regions of both higher (Galactic plane, Orion region, Ophiuchus
complex, LMC, ...) and lower (Ecliptic pole regions) fluctuation levels can be
clearly discerned.

Our final sample is almost 100\% complete over the unmasked BSS area above 2 Jy
(only 1 source with $S_{\rm BSS}=2.06$ Jy is detected with $\hbox{SNR} <5$
level). The completeness above 1 Jy of the SB sample is 89\%, and that of the
IB is 80\%. It increases respectively to 91\% and 82\% including the non-blind
detections.

\subsection{Simple blind vs iterative blind approach}

The performances of SB and IB methods are similar. The SB method recovers with
$\hbox{SNR} >5$ more sources also in highly contaminated regions, but in those
regions the fraction of spurious detections is higher. Hence, choosing among
the two approaches amounts to choosing among slightly higher completeness (SB)
and slightly higher reliability (IB). On the whole, there is no clear advantage
in adopting the more complex IB approach, and we will no longer consider it.

\section{Blind and non-blind source detection on all-sky WMAP 5-yr
maps.}\label{sec:newps5yr}

\begin{table*}
\caption{Summary of the properties of the NEWPS\_5yr\_$5\sigma$
catalogue. Areas with $\sigma_{\rm pixel} > 1.5\sigma_{\rm
median}$ have been left aside. }
\label{tab:sample}
 \begin{minipage}{200mm}
  \begin{tabular}{lccccc}
  \hline
&\textbf{Total}  &23 GHz    & 33 GHz    & 41 GHz    & 61 GHz    \\
  \hline

$\sigma_{\rm median}$ (mJy)                               & & 169       &     206   &  196      &   250     \\
Area selected [sr]                                        & &10.66      & 11.12     & 11.24     & 11.25     \\
Simple blind detections                                   &&405        &281        &275        & 147       \\
Additional non-blind detections                           &&28& 26        & 26        & 14        \\
\textbf{Total number of objects $|b|>5^\circ$ } &         516     & 433       &      307  &      301  &       161 \\
\textbf{Total number of objects with $5<|b|<10^\circ$ }     &51&&           &           &           \\
Total number of objects within the LMC boundaries         &   10&&           &           &           \\
\textbf{Total number of identified Galactic objects }     &27  &11        &14         &21        &10         \\
\textbf{Total number of identified extragalactic objects }& 457 &406        &281        &268        &147        \\
\textbf{Total number of objects missing a consistent counterpart}&32 & 16          &12         &12         &4          \\
Number of sources in WMAP\_5yr                            &   352&&           &           &           \\
Number of sources only in WMAP\_5yr                        &36            &           &           &           &           \\
 \hline
\end{tabular}
\end{minipage}
\end{table*}

\begin{table*}
\caption{Sample of the NEWPS\_5yr\_$5\sigma$  catalogue.
`$r_{xx}$' is the ratio between the $\sigma_{\rm pixel}$
at the source and the median $\sigma_{\rm median}$ in the
WMAP band at $xx$ GHz. The full table is available on the webpage
http://max.ifca.unican.es/caniego/NEWPS/} \label{tab:NEWPS5yr}
 \begin{minipage}{250mm}
 \setlength{\tabcolsep}{0.5mm}
  \begin{tabular}{rrrrrrrrrrrrrrrrrlll}
  \hline
    \# & RA     &  $\delta$ &   l      &      b  & S$_{23}$ & S$_{33}$&
S$_{41}$& S$_{61}$ & $\sigma_{23}$& $\sigma_{33}$& $\sigma_{41}$&
$\sigma_{61}$& r$_{23}$& r$_{33}$& r$_{41}$& r$_{61}$& flags& z& Id.\\
  \hline
     1  &    0.9780&   68.6173 & 118.6020 &   6.1460 &         . &           .&           . &        2141 &      . &           . &           . &         250& .&      .&      .&      1.35&...BG..u&      .& ...\\
     2  &    1.0589&  -47.6401 & 324.0430 & -67.5070 &       947 &         967&           . &           . &    159 &         174 &           . &           .& 0.95&   0.91&      .&      .&BN....We&      .& PKS 0002-478   \\
     3  &    1.5432&   -6.3719 &  93.4960 & -66.6210 &      2523 &        2672&        2604 &        2328 &    170 &         224 &         200 &         257& 1.01&   1.04&   1.01&   1.02&BBBB..We& 0.3470& PKS 0003-066   \\
     4  &    2.6655&   11.0522 & 107.0640 & -50.5590 &      1253 &        1235&        1274 &           . &    163 &         217 &         198 &           .& 0.98&   1.03&   1.02&      .&BBB...We& 0.0893& PKS J0010+1058 \\
     5  &    3.2328&  -39.9448 & 332.4160 & -74.9010 &      1182 &        1385&         970 &        1144 &    153 &         190 &         185 &         227& 0.91&   0.92&   0.94&   ..95&BBBN..We&      .& PKS 0010-401   \\
     6  &    4.8773&   20.2997 & 112.8470 & -41.9460 &       974 &           .&           . &           . &    165 &           . &           . &           .& 0.98&      .&      .&      .&B.....We&      .& PKS 0017+200   \\
     7  &    4.9167&   26.0203 & 114.0680 & -36.3040 &      1140 &           .&           . &           . &    163 &           . &           . &           .& 0.97&      .&      .&      .&B.....We& 0.2840& PKS J0019+2602 \\
     8  &    4.9285&   73.5179 & 120.6480 &  10.7870 &         . &           .&        1469 &           . &      . &           . &         254 &           .& .&      .&   1.33&         .&..N....e& 1.7810& GB6 J0019+7327 \\
     9  &    6.5216&  -26.1011 &  41.7750 & -84.2420 &       904 &           .&           . &           . &    157 &           . &           . &           .& 0.94&      .&      .&      .&B.....We& 0.3220& PKS 0023-26    \\
    10  &    6.6014&  -35.1350 & 335.1410 & -80.3720 &      1299 &        1190&        1505 &        1443 &    153 &         190 &         185 &         238& 0.92&   0.95&   0.94&   0.95&BBBB..We&      .& PMN J0026-3512 \\

 \hline
\end{tabular}
\end{minipage}
{Note: Coordinates in degrees, flux densities in mJy. Flags column: the
first four characters indicate for each WMAP channel used whether the flux
density was estimated with the blind (B) or the non-blind (N) method; the fifth
character identifies sources at $|b|<10^\circ$ (G); the sixth character
indicates sources within $5^\circ$ from the center of the Magellanic Cloud
(L); a `W' as the seventh character stands for sources in the
WMAP\_5yr catalogue; the eighth character gives the source
identification: `g' is for Galactic, `e' is for extragalactic sources, and `u'
is for sources that miss a consistent counterpart.\hfill}
\end{table*}

\begin{figure}
 \includegraphics[width=6cm,angle=90]{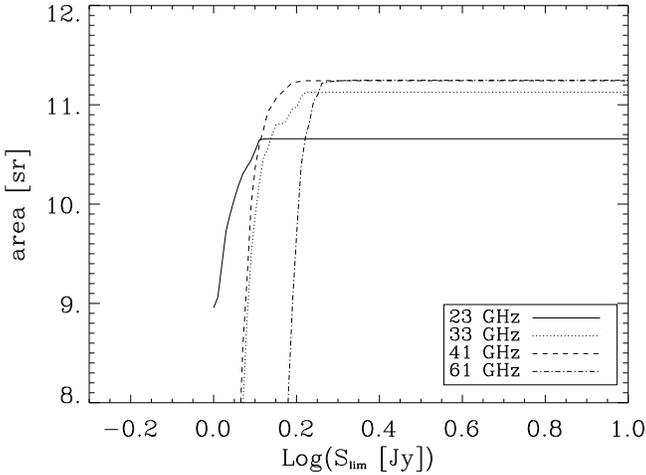}\\
\caption{Effective area as a function of the flux limit $S_{\rm
lim}= 5\sigma_{\rm pixel}$, for the simple blind method. }\label{fig:corrfact}
\end{figure}

\begin{figure*}
 \includegraphics[width=12cm,angle=90]{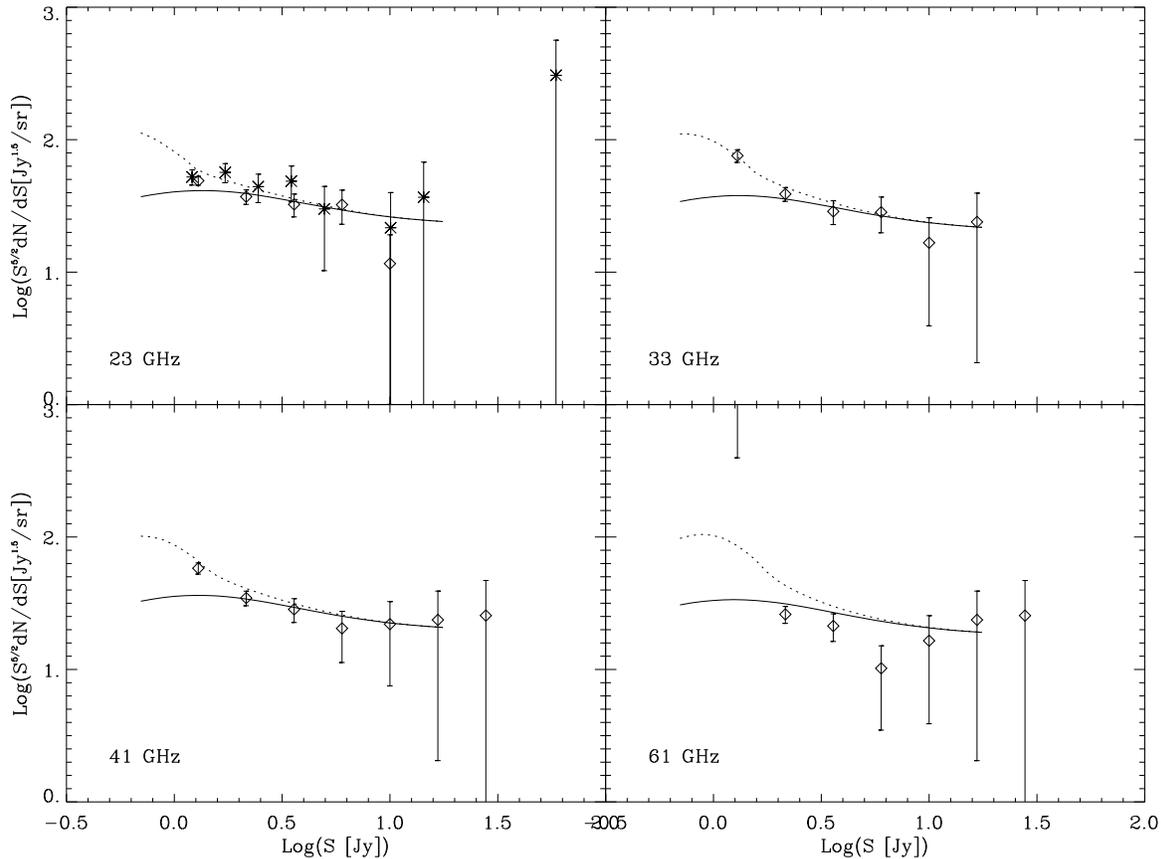}\\
\caption{Differential WMAP counts, normalized to $S_{\rm Jy}^{-2.5}$, estimated
from the WMAP data (diamonds). The 23 GHz counts are compared with the ATCA 20
GHz ones (asterisks). The solid lines show the predictions of the model by De
Zotti et al. (2005). The dotted lines illustrate the effect of the Eddington
bias by showing the model counts convolved with a Gaussian error distribution
with $\sigma = 0.34$, 0.42, 0.4, 0.5 Jy at 23, 33, 41, and 61 GHz,
respectively. The value of $\sigma$ at 23 GHz was obtained by comparison with
the BSS measurements. At higher frequencies we assumed that the true errors on
flux measurements are twice the median errors yielded by the simple blind
approach, as found at 23 GHz. The convolution has been computed integrating
down to a minimum flux equal to $S/10$.}\label{fig:counts}
\end{figure*}

The analysis of the BSS sample, described above, provides useful
guidance for the analysis of all-sky WMAP 5-yr maps at 23, 33, 41
and 61 GHz, to produce the NEWPS\_5yr catalogue. For each map we
have performed the following steps.

\begin{itemize}
\item We have carried out a simple blind search over the whole sky.
\item We have produced maps of the mean noise values per pixel
corresponding to $\hbox{N}_{\rm side}=32$, $\sigma_{\rm pixel}$,
and computed the median of such values for $|b|<5^\circ$,
$\sigma_{\rm median}$.
\item We have masked all the pixels with noise level $>
1.5\sigma_{\rm median}$, which, as found in the previous analysis, contain
a large fraction of the doubtful detections.

\item We have taken all the sources with $\hbox{SNR}>5$ outside the masked area as true
detections.

\item By counting the number of pixels for which $5\sigma_{\rm
pixel}$ is smaller than any given flux density limit, $S_{\rm lim}$, we
obtain the effective area, as a function of $S_{\rm lim}$, to be used to
estimate the differential source counts (see Fig.~\ref{fig:corrfact}). The
decrease of the effective area with decreasing flux density is consistent
with the decrease of the fraction of detected BSS sources reported in
\S\,\ref{completeness}. The maximum effective area is given, for each WMAP
channel, in the second row of Table~\ref{tab:sample}.
\end{itemize}

Next, we carried out a non-blind search on the 5-yr WMAP maps at
the positions of sources in the NEWPS\_3yr\_$3\sigma$  sample (see
\S\,\ref{sec:intro}). This search has produced 28 additional
$\hbox{SNR}>5$ detections at 23 GHz.

The main properties of the NEWPS\_5yr\_$5\sigma$ catalogue,
including $\hbox{SNR}>5$ detections obtained with both the blind
and the non-blind approach, are summarized in
Table~\ref{tab:sample}. The sample totals 516 sources detected in
the regions where $\sigma_{\rm pixel}\le 1.5 \sigma_{\rm median}$.
A search in the NED yielded 457 identifications with extragalactic
sources and 27 identifications with Galactic objects. Only for 32
objects no consistent counterparts were found (5 of them are
clearly in the region of the nebula NGC 1499). Even if they are
all spurious, the reliability of our sample is of 93.8\%, close to
that found from the comparison with the BSS sample.

Of the 388 WMAP\_5yr sources in the sky region covered by the
NEWPS\_5yr\_$5\sigma$ catalogue, 352 have been recovered. All the other 36 have
detections below our $\hbox{SNR}=5$ threshold. On the other hand, the
NEWPS\_5yr\_$5\sigma$ catalogue contains 164 objects not in WMAP\_5yr. 31 of
the new sources do not have consistent counterparts in low frequency catalogs
and may therefore be spurious.

Of the 64 sources detected by Chen \& Wright (2008) in the 3-yr
catalogue, 50 are in our NEWPS\_3yr at 61 GHz. All the 64 objects
have been recovered in the present analysis, but 6 of them are
below the $\hbox{SNR}=5$ threshold. Our NEWPS\_5yr\_$5\sigma$
catalogue also includes all the sources detected by Nie \& Zhang
(2007) outside the LMC region and at $|b|>5^\circ$, not present in
the 3-year WMAP catalogue.

The first lines of the NEWPS\_5yr\_$5\sigma$ catalogue are shown
in Table~\ref{tab:NEWPS5yr}. The full catalogue, the catalogue of
the 3 SNR detections, and the noise maps are available on the web
page http://max.ifca.unican.es/caniego/NEWPS/.

\begin{table}
\caption{The differential normalized source counts ($\log(S^{5/2}
dN/dS[\hbox{Jy}^{1.5}/\hbox{sr}])$) of WMAP sources for each channel. No
correction for the Eddington bias has been applied (see the text and the
caption of Fig.~\protect\ref{fig:counts} for details).} \label{tab:counts}
 \begin{minipage}{180mm}
 \setlength{\tabcolsep}{1.4mm}
  \begin{tabular}{ccccc}
  \hline
   $\log\ $S [Jy]    &  23 GHz &  33 GHz  &41 GHz  &61 GHz \\
  \hline
   0.1 &  $1.69^{+0.03}_{-0.03}$ & $  1.87^{+0.04}_{-0.05}$ & $  1.77^{0.04}_{-0.04} $&$3.51^{0.27}_{-0.91} $\\
   0.3 &  $1.57^{+0.05}_{-0.06}$ & $  1.59^{+0.05}_{-0.06}$ & $  1.54^{0.05}_{-0.06} $&$1.42^{0.06}_{-0.07} $\\
   0.5 &  $1.51^{+0.08}_{-0.09}$ & $  1.46^{+0.08}_{-0.10}$ & $  1.45^{0.08}_{-0.10} $&$1.33^{0.09}_{-0.12} $\\
   0.8 &  $1.51^{+0.11}_{-0.15}$ & $  1.45^{+0.11}_{-0.16}$ & $  1.31^{0.13}_{-0.26} $&$1.01^{0.17}_{-0.47} $\\
   1.0 &  $1.06^{+0.22}_{-1.06}$ & $  1.22^{+0.19}_{-0.63}$ & $  1.34^{0.17}_{-0.47} $&$1.22^{0.19}_{-0.63} $\\
   1.2 &                         & $  1.38^{+0.22}_{-1.06}$ & $  1.37^{0.22}_{-1.06} $&$1.37^{0.22}_{-1.06} $\\
   1.4 &                         &                          & $  1.41^{0.26}_{-11.4}$&$1.41^{0.27}_{-11.4}$\\
 \hline
\end{tabular}
\end{minipage}
\end{table}

The counts of WMAP sources for each channel are presented in
Table~\ref{tab:counts} and in Fig.~\ref{fig:counts}. They have been estimated
calculating the effective area over which each source could have been detected
(Fig.~\ref{fig:corrfact}) and summing the inverse areas in the flux density bin
of interest (Katgert et al. 1973). Error estimates use the approximation
formulae for a Poisson statistics recommended by Gehrels (1986), with an
effective number of sources
\begin{equation}\label{eq_vvir}
n_{\rm eff}=\frac{\left(\sum_i (1/A_i)\right)^2}{\sum_i (1/A_i)^2}\ .
\end{equation}
As expected, the counts are systematically overestimated at the
faintest flux densities, by effect of the Eddington bias. At 23
GHz, the De Zotti et al. (2005) model suggests that the
overestimate is of about 15\% at 2 Jy, and rapidly increases with
decreasing flux (it is $\simeq 30\%$ at 1.5 Jy, and reaches a
factor of almost 2 at 1 Jy). 

\section{Discussion and conclusions}\label{sec:Conclusions}

We have analyzed the efficiency in source detection and flux
density estimation of \emph{blind} and \emph{non-blind} detection
techniques based on the MHW2 filter applied to the WMAP 5-year
maps. Comparing with a complete sample of radio sources, the AT20G
Bright Source Sample (BSS; M08), selected at 20 GHz, close to the
lowest WMAP frequency, with very high signal-to-noise flux
measurements, and almost contemporary to the WMAP survey, we
estimated the completeness, the reliability, and the accuracy of
flux density and error estimates for the samples detected with the
two approaches.

We found that flux density estimates are essentially unbiased except at the
faintest flux densities ($S_{\rm BSS}< 840\,$mJy), where the fraction of the
source intensity peaks amplified by positive fluctuations due to other
components (Galaxy, CMB, noise) within the WMAP beam becomes substantial and
the source counts are correspondingly overestimated. This is a manifestation of
the Eddington bias, enhanced by the fact that the true errors on flux density
estimates turn out to be about a factor of 2 higher than the errors estimated
by our procedure (see \S\,\ref{sec:acc}). The difference is due to the
filtering of the maps that increases the signal-to-noise ratio by smoothing the
fluctuation field. No clear-cut criterion capable of identifying sources
affected by this problem using only WMAP data was found. However, coupling the
estimate of the true uncertainties on source fluxes, obtained by comparison
with the high signal-to-noise ATCA 20 GHz measurements, with information on
counts below the WMAP detection limit we may estimate the corresponding
corrections on source counts. Using the De Zotti et al. (2005) model, that
incorporates  the information on counts provided by the AT20G (M08; Ricci et
al. 2004) and 9C (Waldram et al. 2003) surveys at nearby frequencies we
estimate that, in the K-band, the counts have to be corrected downward by about
15\% at 2 Jy, and by a factor of almost 2 at 1 Jy.

At higher flux densities most (17 out of 19) of probably spurious detections
are at relatively low Galactic latitudes ($|b|<20^\circ$), suggesting that the
observed intensity peaks are largely due to small scale structure in the
Galactic emission. Excluding the areas where the rms fluctuations are more than
50\% higher than the $|b|>5^\circ$ median approximately halves the number of
dubious candidates, at a modest cost ($\simeq 7$--10\%) in terms of useful
area. If all dubious sources are spurious, the reliability of the sample is
95.5\%.

The blind detection approach applied to the all-sky WMAP maps,
excluding the Galactic plane region ($|b|< 5^\circ$) and the areas
where the rms fluctuations are more than 50\% higher than the
median value at $|b|< 5^\circ$, has found 488 candidate sources
with $\hbox{SNR}>5$ in at least one WMAP channel. The non-blind
approach has added 28 further objects, raising the total to 516,
to be compared with the 388 sources listed in the WMAP 5-yr
catalog (Wright et al. 2008). Almost all (484) sources in our
sample were previously catalogued extragalactic (457) or Galactic
(27) objects. The remaining 32 candidate sources do not have
counterparts in lower frequency all sky surveys with comparable
flux densities and may therefore be just high peaks in the
distribution of other components present in the maps. If they are
all spurious, the reliability of the sample is 93.8\%.

\section*{Acknowledgements}

Partial financial support for this research has been provided to
MM, JGN, and GDZ by the Italian ASI (contracts Planck LFI Activity
of Phase E2 and I/016/07/0 `COFIS') and MUR, and to JLS by the
Spanish MEC. MLC acknowledges a postdoctoral fellowship from the
Spanish MEC. JGN acknowledges a postdoctoral position at the
SISSA-ISAS (Trieste). JLS thanks the CNR ISTI in Pisa for their
hospitality during his sabbatical leave. This research has made
use of the NASA/IPAC Extragalactic Database (NED) which is
operated by the Jet Propulsion Laboratory, California Institute of
Technology, under contract with the National Aeronautics and Space
Administration. Some of the results in this paper have been
derived using the HEALPix (G\'orski et al., 2005) package. Part of
this analysis has been carried out using Grid infrastructure in
the framework of the project EGEE (reference FP7 INFSO-RI 222667).

\bsp

\label{lastpage}

\end{document}